\documentstyle[12pt]{article}
\begin{document}

\baselineskip 24pt

\newcommand{\sheptitle}
{Low Scale Technicolour at LEP}

\newcommand{\shepauthor}
{S. F. King,}

\newcommand{\shepaddress}
{Physics Department,\\University of Southampton,\\Southampton,\\SO9 5NH,\\U.K.}

\newcommand{\shepabstract}
{We discuss the phenomenology of an
$SU(2)_{TC}$ technicolour model
with a low technicolour confinement scale ${\Lambda}_{TC} \sim 50-100 GeV$.
Such a low technicolour scale may give rise to
the first hints of technicolour being seen at LEPI and
spectacular technicolour signals at LEPII.}

\begin{titlepage}
\hfill SHEP 92/93-22
\vspace{.4in}
\begin{center}
{\Huge{\bf \sheptitle}}
\bigskip \\ \shepauthor \\ {\it \shepaddress} \\ \vspace{.5in}
{\bf Abstract} \bigskip \end{center} \setcounter{page}{0}
\shepabstract
\end{titlepage}

\newcommand{\beq}{\begin{equation}}
\newcommand{\eeq}{\end{equation}}
\newcommand{\beqarr}{\begin{eqnarray}}
\newcommand{\eeqarr}{\end{eqnarray}}
\newcommand{\psibarpsi}{\mbox{$<\bar{\psi}\psi>_{M_{ETC}}$}}
\newcommand{\alphaetc}{\mbox{$\alpha_{ETC}$}}

\newpage

\setcounter{page}{1}
\pagestyle{plain}

Recently we proposed a dynamical model of leptons in which the
leptons were endowed with three colours (red, white and blue) where
two of the colours (red and blue)
split off to form a technicolour (TC) group $SU(2)_{TC}$,
while the third colour (white) is identified with the physical leptons
\cite{4}. Fermion masses were generated by
fourth family condensates \cite{4}.
In this paper we discuss the phenomenology of
a three family version of this model
on the assumption that the TC group confines at a low technicolour
confinement
\footnote{Our definition of the confinement scale is that it
is equal to one half of the mass of the lowest lying
vector resonance.}
scale ${\Lambda}_{TC} \sim 50-100 GeV$,
which is a natural assumption in this model.
Such a low technicolour scale may give rise to
the first hints of technicolour being seen at LEPI and
spectacular technicolour signals at LEPII.

When the three lepton colours have
split apart as discussed above,
the model corresponds to the moose diagram in Fig.1.
The exchange of the various heavy gauge bosons which mix
together via loops of preons gives rise to contact operators
in the effective low energy theory which describes the
three families of leptons
and technifermions.
The key features of this model relevant to our phenomenological
discussion are
listed below:
\begin{itemize}
\item
Technicolour $SU(2)_{TC}$ confines at $\Lambda_{TC}\sim 50-100\ GeV$.
\item
There are three technidoublets closely associated with three lepton
families, but only coupling to them weakly via chirality conserving operators.
\item
One dominant technidoublet $T$ (associated with the $\tau$ family)
is mainly responsible for
electroweak symmetry breaking
and has a heavy dynamical mass on the order of $500-1000 GeV$.
\item
The remaining two technidoublets $t_1,t_2$
(associated with the $e$ and $\mu$ families)
have light dynamical masses on the order of ${\Lambda}_{TC} \sim 50-100 GeV$.
\end{itemize}
The following discussion will apply to any TC model which has the above
features, and not just the model in Fig.1.

To begin with let us consider the dominant
technidoublet $T=(P,M)$
\footnote
{Each techniquark carries TC but not ordinary colour, and the left-handed
techniquarks $T_L$ form an $SU(2)_L$ doublet just like ordinary quarks
but with weak hypercharge $Y=0$. The right-handed techniquarks $P_R,M_R$
have $Y=1/2,-1/2$, respectively.
The electric charge generator is given by $Q=T_{L3} + Y$ so that
the plus ($P$) and minus ($M$) techniquarks have charges given by
$Q=\pm 1/2$,respectively.}.
In this model the exchange of heavy flavour gauge bosons
induces operators of the form \cite{4}
$G_T(\bar{T}_{L}T_{R})(\bar{T}_{R}T_{L})$.
There are similar contact terms
of the form $G_{\tau}(\bar{\tau}_{L}\tau_{R})(\bar{\tau}_{R}\tau_{L})$
which induce a
tau lepton condensate \cite{4} and other contact terms
such as
$G_{\tau \mu}(\bar{\tau}_{L}\tau_{R})(\bar{\mu}_{R}\mu_{L})$
allow the tau mass to be fed down to the muon (and similarly
for the electron). Other contact terms such as
$G_{\mu}(\bar{\mu}_{L}\mu_{R})(\bar{\mu}_{R}\mu_{L})$
have smaller coefficients.
Quark masses are generated by top and bottom
quark condensates \cite{5,6}.
Since we assume a tau condensate then $G_T\approx G_{\tau}$
must therefore be strong, leading to a condensate
$<\bar{P}P+\bar{M}M>\neq 0$.
If the operator respects isospin then the
pattern of symmetry breaking
\footnote{The full $SU(4)$ symmetry of a one-doublet
$SU(2)_{TC}$ model is broken here
by the four-technifermion operators. We assume these operators
respect isospin symmetry
for convenience only; in reality there will be some isospin violation.}
expected is just
\beq
SU(2)_L\otimes SU(2)_R\rightarrow SU(2)_{L+R}
\eeq
yielding a triplet of technipions
$\Pi_{TC}^{\pm ,0}\sim \bar{T}\sigma^{\pm ,3}\gamma_5T$,
where $\sigma^a$ are the Pauli matrices.
In such a theory the technipion decay constant may be much larger
than $\Lambda_{TC}$ since chiral symmetry breaking is driven
mainly by the above contact operator \cite{3}.
Thus we may have,
$<0|J^a_{\mu 5}|\Pi_{TC}^b>=iF_{TC}q_{\mu}\delta^{ab}$,
where $F_{TC}\sim 245 GeV$, and the current is
$J^a_{\mu 5}=\bar{T}\gamma_{\mu}\gamma_5\sigma^aT$.
All these technipions get eaten,
and the remaining technihadrons have
masses set by the enhanced dynamical masses of the technifermions,
and are in the LHC/SSC range.

Now let us extend our discussion to include one of the other
technidoublets $t_2$ in this model, where $t_2$ is associated
with the muon family. The technidoublet $t_1$ will be considered later.
There are now two technidoublets $T=(P,M),\ t=(p,m)$, which have identical
quantum numbers (we have dropped the subscript on $t_2$ for convenience).
The set of operators in this case are
of the form \cite{4}
\beqarr
G_T(\bar{T}_{L}T_{R})(\bar{T}_{R}T_{L}),\
G_t(\bar{t}_{L}t_{R})(\bar{t}_{R}t_{L}),\
G_{tT}(\bar{T}_{L}T_{R})(\bar{t}_{R}t_{L}).
\eeqarr
where $G_t,G_{tT}\ll G_T$ since $G_t\approx G_{\mu}$ and
$G_{tT}\approx G_{\tau \mu}$.
In this case the theory naturally splits into two parts,
a high scale TC sector associated with the technifermions $T$
which form condensates driven by the operators
associated with the scale $\Lambda \sim 1 TeV$,
as discussed above,
and a low scale TC sector associated with technifermions $t$
which form condensates driven by TC interactions at the scale
$\Lambda_{TC}\sim 50-100 \ GeV$.
The low scale TC sector
again has an approximate global symmetry as in Eq.(1),
but now there is a vacuum alignment problem
which depends on the relative strength of the contact term
and the TC gauge forces. The TC gauge forces tend to favour the
chiraly invariant condensate
$<\bar{t^c_L}t_L+\bar{t^c_R}t_R>\neq 0$,
while the contact terms prefer the chiral symmetry
breaking condensates,
$<\bar{t_L}t_R+\bar{t_R}t_L>\neq 0$.
We shall return to this point later.
For now we shall assume the latter condensates form,
yielding a triplet of technipions
$\pi_{TC}^{\pm ,0}\sim \bar{t}\sigma^{\pm ,3}\gamma_5t$, with
$<0|j^a_{\mu 5}|\pi_{TC}^b>=if_{TC}q_{\mu}\delta^{ab}$,
where $f_{TC}\sim 10-25 GeV$, and the current is
$j^a_{\mu 5}=\bar{t}\gamma_{\mu}\gamma_5\sigma^at$.
There are two points of interest concerning the low scale TC sector.
Firstly the low scale technipion triplet do not get eaten.
\footnote
{Strictly the eaten and physical technipions are given by
$| eaten \; technipion>  =
 \frac{F_{TC}|\Pi_{TC}^{\pm ,0} >+f_{TC}| \pi_{TC}^{\pm ,0}>}
{\sqrt{F_{TC}^2 + f_{TC}^2}}$ and
$| physical \;  technipion>  =
 \frac{f_{TC}|\Pi_{TC}^{\pm ,0} >-F_{TC}| \pi_{TC}^{\pm ,0}>}
{\sqrt{F_{TC}^2 + f_{TC}^2}}$.
The full current is $J_{5\mu}+j_{5\mu}$,
where,
$<0|J_{5\mu}+j_{5\mu} |\Pi_{TC}^{\pm ,0}>=F_{TC}q^{\mu}$
and $<0|J_{5\mu}+j_{5\mu}|\pi_{TC}^{\pm ,0}>=f_{TC}q^{\mu}$,
which implies,
$<0|J_{5\mu}+j_{5\mu}|eaten \ technipion>={\sqrt{F_{TC}^2 + f_{TC}^2}}
     q^{\mu}$
and $<0|J_{5\mu}+j_{5\mu}|physical \ technipion>=0$.}
More importantly the low scale TC sector may be accessible to existing
colliders such as LEP, as we now discuss.

The physical technipions $\pi_{TC}^{\pm,0}$ will receive a mass from
the mixed operator with coefficient $G_{tT}$ in Eq.(2).
The ``explicit'' masses of the
low scale technifermions resulting from this operator
are obviously just $m_{p,m}=G_{tT}<(\bar{T}_{L}T_{R})>$,
which may be estimated since $G_{tT}\approx G_{\tau \mu}$.
These masses break the chiral symmetry of the second
technidoublet resulting in a physical technipion mass
analagous to the way in which the physical pion mass
results from explicit quark masses.
The technipion mass $m_{\pi_{TC}}$ may be estimated by scaling up the
usual result for the ordinary pion mass $m_{\pi}$,
\beq
m_{\pi_{TC}}= m_{\pi}\sqrt{\left(\frac{m_p + m_m}{m_u + m_d}
\right)\frac{f_{TC}}{f_{\pi}}}.
\eeq
Using Eq.(3) combined with our estimate of $m_{p,m}$
we find rather heavy technipion masses of order tens of GeV.
LEPI sets a limit on the mass of charged technipion masses of around $M_Z/2$.

Assuming the technifermions have very small
couplings to leptons
(the chirality conserving couplings are discussed later)
the charged technipions $\pi_{TC}^{\pm}$ will decay
predominantly via virtual W exchange, analagous to ordinary charged
pion decay. Thus for example the width into leptons is given by,
\beq
\Gamma (\pi_{TC}^{\pm}\rightarrow l^{\pm}\nu_l)
= \frac{f_{TC}^2}{f_{\pi}^2}\frac{m_l^2}{m_{\mu}^2}
\frac{m_{\pi_{TC}}}{m_{\pi}}\frac{1}{(1-\frac{m_{\mu}^2}{m_{\pi}^2})^2}
\Gamma (\pi^{\pm}\rightarrow \mu^{\pm}\nu_{\mu}).
\eeq
The largest such decay channels are thus $bc,cs,\tau \nu_{\tau}$.
The neutral technipions $\pi_{TC}^0$ may also decay by this mechanism,
via virtual Z exchange,
with the largest partial width,
\beq
\Gamma (\pi^{0}_{TC}\rightarrow \bar{b}b)
\approx
3\left(\frac{m_b^2}{m_l^2}\right)
\Gamma (\pi^{\pm}_{TC}\rightarrow l^{\pm}\nu_l).
\eeq
$\pi_{TC}^0$ will also decay into two photons via a chiral symmetry
suppressed anomalous $\pi_{TC}^0\gamma \gamma$ coupling.
The $\pi_{TC}^0\gamma \gamma$  coupling proportional to
\beq
S_{\pi_{TC}^0\gamma \gamma}
=N_{TC}Tr[\frac{\sigma^3}{2}
(\{\Gamma_V,\Gamma_V\}+\{\Gamma_A,\Gamma_A\})\frac{\mu M}{\Lambda_{TC}^2}]
\eeq
where $\Gamma_{A}=0$, $\Gamma_{V}=Q=e\frac{\sigma_3}{2}$,
$M=diag(m_p,m_m)$ is the technifermion mass matrix,
and $Tr \mu M \approx m_{\pi_{TC}}^2$.
The usual anomalous coupling without the mass factor is zero in this model.
Providing that there is some isospin
violation (i.e. $m_p\neq m_m$) the contribution to this coupling
will be non-zero, and will be only mildly suppressed relative to the
usual anomalous coupling
by a factor of roughly $m_{\pi_{TC}}^2/\Lambda_{TC}^2$.
The partial width is given by \cite{8},
\beq
\Gamma (\pi_{TC}^0 \rightarrow \gamma \gamma)=
A_{\pi_{TC}^0\gamma \gamma}^2
\left( \frac{\alpha^2}{16\pi^3f_{TC}^2}\right)m_{\pi_{TC}^0}^3
\eeq
where $S_{\pi_{TC}^0\gamma \gamma}=A_{\pi_{TC}^0\gamma \gamma}2e^2$,
where in the present model
$A_{\pi_{TC}^0\gamma \gamma}=-(1/8)\left(\frac{m_{\pi_{TC}^0}}
      {\Lambda_{TC}}\right)^2$.
Despite its small partial width, this will be an important
decay mode of the $\pi_{TC}^0$.

Apart from the low-scale technipions, the technidoublet $t$
will give rise to technivector mesons $V$
analagous to the QCD vector resonances.
However here the masses of such technivectors will be
{\em an order of magnitude} smaller than in conventional TC.
For example we may expect a $J^{PC}=1^{--}$
technirho triplet ${\rho}_{TC}^{\pm ,0}$ and techniomega singlet
${\omega}_{TC}$ with masses in the LEPII range 100-200 GeV.
The vector masses may be estimated by scaling up
the ordinary $\rho$ and $\omega$ mass
$m_{\rho_{TC} ,\omega_{TC}}\approx m_{\rho ,\omega}\frac{f_{TC}}{f_{\pi}}$.
The technidoublet $t$ has photon and Z couplings,
\beq
A_{\mu}\bar{t}\gamma^{\mu}Qt
+Z_{\mu}\bar{t}\gamma^{\mu}(\Gamma_V +\Gamma_A\gamma_5)t
\eeq
where
\beq
Q=e\frac{\sigma^3}{2}, \ \ \Gamma_V=\frac{e}{\tan 2\theta_w}
\frac{\sigma^3}{2}, \ \
\Gamma_A=\frac{-e}{\sin 2\theta_w}\frac{\sigma^3}{2}.
\eeq
Using vector meson dominance arguments, combined with scaling-up
arguments, we write,
\beq
\bar{t}\gamma^{\mu}
\frac{\sigma^a}{2}t\rightarrow
\frac{m_{{\rho}_{TC}}^2}{g_{{\rho}_{TC}}}{\rho^{a\mu }}_{TC}
\eeq
where $g_{{\rho}_{TC}}\approx g_{\rho}$,and $g_{\rho}=\sqrt{12\pi }$.
Thus the ${\rho}_{TC}^0$ may be detected at LEPII via its couplings to the
photon and Z,
\beq
\frac{m_{{\rho}_{TC}}^2}{g_{{\rho}_{TC}}}
{\rho_{TC}}^{0,\mu}\left[eA_{\mu} +
\frac{e}{\tan 2\theta_w}Z_{\mu}\right],
\eeq
leading to resonances in R.
The techniomega $\omega_{TC}^0$ being associated with the
isosinglet current $\bar{t}\gamma^{\mu}t$ does not couple to the photon or Z.

If the technirho mass is above
two technipion threshold then pairs of technipions may be
produced at resonantly enhanced rates. The partial width of the technirho
into a pair of charged technipions is given by a scaling argument as,
\beq
\Gamma (\rho_{TC}\rightarrow \pi^+_{TC} \pi^-_{TC})
=\frac{m_{\rho_{TC}}}{m_{\rho}}
\frac{\beta_{\pi_{TC}}^3}{\beta_{\pi}^3}\Gamma (\rho \rightarrow \pi^+ \pi^-)
\eeq
where $\beta_{\pi (TC)}=(1-\frac{4m_{\pi (TC) }^2}{m_{\rho (TC) }^2})^{1/2}$.
The technivector mesons may also have direct couplings to leptons
via four-fermion operators which couple technifermions to leptons
but do not violate chirality, which arise from exchange
of the flavour diagonal gauge bosons. Neutrino couplings may be
controlled independently in this model.
In addition $t=(p,m)$ (where $m$ is effectively a coloured muon) will have
independent couplings to muons arising from heavy lepton colour
gauge bosons which convert a lepton into a technicoloured
lepton. Thus direct techniomega coupling to leptons
may originate from four-fermion operators of the form
$H_l \bar{t}\gamma_{\mu}t\bar{l}\gamma_{\mu}l$
where $l=e,\mu,\tau ,\nu_e ,\nu_{\mu} ,\nu_{\tau}$.
The relative strengths of
each of the neutrino and muon couplings
can be adjusted in this model.
The operator induces a direct lepton coupling
\footnote{The same operator
will also induce direct couplings of $\pi_{TC}^0$ to leptons
of the form $H_lf_{TC}\partial_{\mu}\pi_{TC}^0\bar{l}\gamma_{\mu}l$.},
\beq
H_l\frac{m_{\omega_{TC}}^2}{g_{\omega_{TC}}}\omega^{\mu}_{TC}
     \bar{l}\gamma_{\mu}l.
\eeq

Recently the L3 collaboration at LEP \cite{9}
reported four events, one $e^+e^-\gamma \gamma$
and three $\mu^+\mu^-\gamma \gamma$,
each with two energetic photons with an invariant mass
$M_{\gamma \gamma}\approx 60 GeV$. Other LEP groups have informally
reported one or two similar events, plus one or two
$\bar{\nu}\nu \gamma \gamma$ and $\bar{q}q \gamma \gamma$
events but such events appear to be consistent with background.
Such events are also consistent with a 60 GeV neutral technipion,
as we now discuss. From Eq.(7)
we find $\Gamma (\pi_{TC}^0 \rightarrow \gamma \gamma)=10 keV$
(assuming $\Lambda_{TC}=50GeV$, $f_{TC}=10GeV$).
Similarly from Eqs.(4,5) we find
$\Gamma (\pi_{TC}^0 \rightarrow \bar{b}b)=4keV$.
The two photon decay is probably too narrow to enable the technipion
to be produced by two photon collisions at LEP.
The partial width of the technipion into $e^+e^-$
is heavily suppressed compared to its width into $\bar{b}b$, and
is well below the sensitivity of the recent TRISTAN search \cite{11},
which is sensitive to electronic widths of a few keV.

Now let us consider the production rate of a 60 GeV technipion at LEP.
The conventional rate \cite{8} for
$Z\rightarrow \pi_{TC}^0l^+l^-$
via a virtual $Z$ is much too small to be relevant at LEP.
However since the technivector mesons in this model are light
then there will be an
additional production mechanism for the technipions
via a $ZV\pi_{TC}^0$ vertex
where V is a technivector meson, either a technirho or
a techniomega.
This coupling is much stronger than typical anomalous
couplings since it involves two strong and one electroweak couplings,
rather than two electroweak and one strong coupling.
This coupling combined with Eq.(13) will allow the decay
\footnote
{A similar mechanism was proposed by Bando and Maekawa\cite{10}, except
that these authors assumed that the real on-shell Z will decay into a
real on-shell $\pi_{TC}^0$ plus a {\em virtual scalar} technihadron.}
\beq
Z\rightarrow \pi_{TC}^0 + \omega_{TC}^{\ast},\
\omega_{TC}^{\ast}\rightarrow \bar{l}l
\eeq
The precise rate will depend on the $\omega_{TC}$ mass,
and the strength of the $Z\omega_{TC}\pi_{TC}^0$ vertex,
and is difficult to estimate in this model.
The decay rate may be approximately
scaled on the standard model Higgs h
production rate via a virtual Z,
\beq
\frac{\Gamma (Z\rightarrow \pi_{TC}^0\bar{l}l)}
{\Gamma (Z\rightarrow h\bar{l}l)}
\approx R_{Z\omega_{TC}\pi_{TC}^0}^2R_{\omega_{TC}\bar{l}l}^2
\frac{M_Z^4}{m_{\omega_{TC}}^4}
\eeq
where we have taken ratios of couplings to those in the standard model process.
There is no reason why these $R$ ratios should not be of order unity,
and, since the $\omega_{TC}$ mass is light in this model,
it appears to be possible to account for the 60 GeV two photons events,
accompanied by pairs of electrons,
muons, taus or neutrinos in adjustable ratios.
The technipion may also be
produced in association with a virtual technirho
which couples to $\bar{q}q$ jets via its photon or Z couplings
in Eq.(11).

A low mass techniomega
will contribute to Bhabha scattering
due to the direct coupling to leptons in Eq.(13),
which will generate an effective four-electron contact interaction.
According to current limits on the coefficient of such
contact interactions we require
\beq
\frac{\alpha_{\omega_{TC}ee}}{m_{\omega_{TC}}^2}
< \frac{4\pi^2}{(1TeV)^2}
\eeq
where
$\alpha_{\omega_{TC}ee}=(H_e\frac{m_{\omega_{TC}}^2}{g_{\omega_{TC}}})^2$
from Eq.(13). Clearly if $\alpha_{\omega_{TC}ee}$ is
of electroweak
strength and the techniomega mass is as low as 100 GeV,
as assumed above, then this bound is close to being realised.
Thus at LEPII one may hope to use Bhabha
scattering as a probe of the techniomega, and perhaps even produce
an on-shell techniomega via the direct lepton coupling in Eq.(13).

In the above discussion we considered only $t_2$ (which we wrote as $t$).
Now we must consider the technidoublet $t_1$ which is associated
with the electron family. In this case it will be necessary
to assume that $t_1$ forms
electroweak preserving condensates since otherwise the tau lepton
mass will be fed down to the electron via flavour operators
generated by diagrams involving internal technifermion condensates.
Operators of the form
$K_L(\bar{e}_{L}\gamma_{\nu}\tau_{L})(\bar{m_1}_{L}\gamma^{\nu}{M}_{L})$
and $K_R(\bar{e}_{R}\gamma_{\nu}\tau_{R})(\bar{m_1}_{R}\gamma^{\nu}{M}_{R})$
may be fused together via technifermion condensates
to form operators which (after a Fierz transformation) have the form
$J(\bar{e_L}e_R)(\bar{\tau_{L}}{\tau}_{R})$. This operator
leads to an electron mass of order $(\frac{\Lambda_{TC}}{\Lambda})m_{\tau}$,
which is too large assuming $(\frac{\Lambda_{TC}}{\Lambda})\approx 0.1$.
Such a mass is barely acceptable for the muon. But for the electron
we must prevent such operators from being constructed and the simplest
way to do this is by assuming that $<\bar{m_1}_L{m_1}_R>=0$.
Instead we assume the
invariant condensates
$<\bar{{p^c_1}_L}{m_1}_L-\bar{m^c_1}_L{p_1}_L + L\rightarrow R>\neq 0$.
It is natural to assume that $t_1$ forms invariant condensates
but $t_2$ forms chiral symmetry breaking condensates
since the contact operators which
prefer the broken condensates are weaker for the electron than for the muon.

It is possible that $t_1,t_2$ both
form chirally invariant condensates
so that all the technihadrons will have masses $\approx 2\Lambda_{TC}$.
In this case it is possible
that the tau lepton mass is generated not from
tau lepton condensates but is fed down from
the TC sector via exchange of the heavy gauge bosons which unify
the three lepton colours \cite{4}.
The contact operators of interest are, after a Fierz transformation,
\beq
K[(\bar{\tau}_{L}\tau_{R})(\bar{M}_{R}M_{L})
+ (\bar{\mu}_{L}\mu_{R})(\bar{m_2}_{R}{m_2}_{L})
+ (\bar{e}_{L}e_{R})(\bar{m_1}_{R}{m_1}_{L})].
\eeq
The tau mass from this mechanism is $m_{\tau}=K<\bar{M}_{R}M_{L}>$,
where K is proportional to an explicit preon mass.
The $e$ and $\mu$ receive no mass from these operators having
assumed invariant condensates for $t_1,t_2$,
but their mass is fed down from the tau lepton in the usual way.
The smallness the tau mass
implies that the chirality breaking operators in Eq.(17)
above are phenomenologically unimportant.

The above mechanism avoids the fine-tuning associated with
tau condensates but
there will still be some fine tuning required
in the quark sector since we require an independent b quark condensate.
All fine tuning can be eliminated by extending this scenario
to include fourth family condensates \cite{4}. In this case
there will be an electroweak breaking
technidoublet $T$ associated with the
fourth lepton family, but it
will be necessary to assume that all three low scale technidoublets
$t_1,t_2,t_3$
(where $t_3$ is now associated with the tau family) form
electroweak preserving condensates in order to prevent
too large a tau mass arising from the operators discussed above.
In such a fourth family variant of this model there
may be chirality violating
couplings as in Eq.(17) controlled by a coupling K which may be much larger
than in the three family scenario. Such couplings will
allow spin zero bound states of mass $\sim 2 \Lambda_{TC}$ to decay
to electrons or muons at
rates which are unsuppressed by the lepton mass.
Such spin zero states may be resonantly produced in $e^+e^-$ collisions.

Finally we must consider the constraints on this model from high precision
tests of the standard model.
At LEPI energies, assuming the $\rho_{TC}$ mass for example
is much greater than $M_Z$
(a somewhat questionable assumption in this model)
the technirho will contribute to
the oblique corrections to the photon and Z propagators.
For example the contribution to the S parameter \cite{7} for
a single technidoublet is estimated to be
$S\approx 0.3\left(\frac{2}{3}\right) \approx 0.2$.
This estimate is based on vector meson dominance and
large $N_{TC}$ scaling arguments in order to estimate the contribution
to dispersion relations. The contribution depends on
scale independent ratios like
$\frac{f_{TC}}{m_{{\rho}_{TC}}}$
so is independent of the fact that the technicolour scale
(and the technirho mass) is low. The dominant technidoublet $T$
may contribute somewhat differently since its dynamics are not QCD-like,
but are controlled by the contact operator.
The technidoublet $t_1$ also forms non QCD-like condensates.
All together from the three technidoublets $T,t_1,t_2$
we might expect a contribution to the S parameter of order 0.6
in this model, subject to the uncertainties mentioned above.
In the four family scenario there will of course be additional
contributions.

In conclusion, we have discussed the LEP phenomenology of a technicolour
model with a low confinement scale. The specific model
is summarised in Fig.1, but our discussion will apply to any pure TC
(i.e. unextended TC) model which has the listed features.
As we have seen such models offer the prospect of exciting
physics at LEPII with perhaps the first signs of new
physics visible at LEPI. The
high scale TC sector may be studied at
LHC or SSC in the usual way. This really is a jam today jam
tomorrow scenario!

{\bf Acknowledgements}

SFK would like to thank Nicholas Evans, Jonathan Flynn,
Lisa Randall, Douglas Ross and Terry Sloan
for discussions and
the SERC for financial support.

\newpage

\begin{center}

{\bf Figure Caption}

{Figure 1}

\end{center}

A three family lepton moose model.
As usual, the circles represent gauge groups and the lines represent
fermions which transform under the gauge groups.
The labelled lines in the lower half of the diagram correspond
to three families of leptons (including right-handed neutrinos) denoted
$(N,E)_L,N_R^c,E_R^c$
and three families of technifermions, denoted $(P,M)_L,P^c_R,M^c_R$.
The three families of leptons and technifermions
share acommon gauge chiral family symmetry
$SU(3)_L\otimes SU(3)_{E}\otimes SU(3)_N$.
The remaining lines represent preons which
transform under the preon gauge groups
$SU(3)_{fE}$ and $SU(3)_{fN}$ which confine at $\Lambda \sim 1 TeV$.
The preons in the upper half of the diagram have explicit
mass matrices, and condense with the preons represented by
horizontal lines at the preon confinement scale.
The preon dynamics breaks the gauged chiral family symmetries
to global chiral family symmetries,
which are explicitly broken by the preon masses,
thereby ensuring a GIM mechanism [2].

\newpage

\bibliographystyle{unsrt}

\end{document}